\documentstyle[12pt]{article}
\begin{document}
\title{Novel interaction induced oscillations in 
correlated electron transport}
\author{F. Claro$^{1}$, J. F. Weisz$^{2}$ and S. Curilef$^{3}$\\
$^{1}$Facultad de F\'{\i }sica, Pontificia Universidad Cat\'{o}lica de\\
Chile, Casilla 306, Santiago 22, Chile\\
$^{2}$CRICYT, C. C. 131, 5500 Mendoza, Argentina\\
$^{3}$Departamento de F\'{i}sica, Universidad Cat\'{o}lica del Norte,\\
Casilla 1280, Antofagasta, Chile}
\date{}
\maketitle

\begin{abstract}
The correlated motion of electrons in a one dimensional system with an
externally applied longitudinal electric field is discussed. Within the
tight binding model we show that in addition to the well known Bloch
oscillations the electron-electron interaction induces time dependent
oscillations of the mobility whose period depends on the strength and range
of the coupling only. The oscillations involve transitions between bands of
bound and unbound states. The case of two electrons is solved in detail and
an extension of the results to more particles is discussed.
\end{abstract}

PACS numbers: 72.30.+q, 73.23.-b, 78.67.Lt

\vskip2pc

There has been much interest recently in nanostructures containing very few
electrons, the entrance or exit of which can be controled by a gate voltage.%
\cite{llsohn} While transport properties in these devices have been well
studied and understood within independent electron approximations or mean
field theories, the effect of correlations is still largely unresolved.\cite
{Ave,Hac} An important question concerns the effect of the electron-electron
interaction on the localization induced by disorder, a subject of recent
controversy.\cite{shep,imry,wein,romm,evan,song,bene,elis,jacq} The usual
approach to deal with this problem is to look for approximate solutions or
estimates of unperturbed properties of the system, such as the decay rate of
the ground state localized wavefunctions. We here take a different route,
and treat explicitely the time dependent response of the ensemble to an
external uniform electric field, in order to judge the ability to conduct
through a study of the mobility. Our main finding is that this quantity
oscillates in time with a period solely determined by the strength and range
of the interaction.

The simplest case to treat is that of just two interacting electrons. We
consider first such a pair, constrained to a one dimensional chain of
lattice parameter $a$, with an electric field F applied along the wire. In
the tight binding model, the amplitudes $C_{l,m}$ for having one particle at
site $l$ and the other at site $m$ at time $t$, obey the equation

\begin{eqnarray}
-\lambda(C_{l+1,m}+C_{l-1,m}+C_{l,m+1}+  \nonumber \\
C_{l,m-1})+E_{l,m}C_{l,m}=i\hbar \frac{dC_{l,m}}{dt},
\end{eqnarray}

\noindent where $E_{l,m}=\epsilon _{l}+\epsilon _{m}+V(l-m)-eFa(l+m)$, with $%
\epsilon _{l}$ the energy at site $l$ and $V(l-m)$ the two body interaction
potential. $\lambda $ is the usual hopping energy. In this model either of
the two charges can hop to its nearest neighbour site. Disorder may be
included by making the site energies {$\epsilon _{l}$} random.

The two electron problem described above is equivalent to that of a single
particle moving in a square lattice with sites on the plane labeled by the
pair $l,m$. The interaction acts like an interface potential, symmetric
about the diagonal $l=m$ where the boundary is located, and the applied
external field is parallel to this line. The spectrum and eigenstates of the
system with no electric field, no disorder and a contact Hubbard interaction 
$V(l-m)=U\delta _{l,m\mbox{ }}$ are easily found for this problem. To obtain
the solution, we make in Eq. (1) the substitution

\begin{eqnarray}  \label{coefficient}
C_{l,m}=e^{i(l+m)ka} f(l-m)
\end{eqnarray}

\noindent in terms of new amplitudes $f$ that depend on the distance $u=l-m$
to the interface only. It is found that $f$ obeys a one-dimensional
equation, with hopping amplitude $-2\lambda \cos (ka)$ and a chain of
defects of local potential $U$ along the line $l=m$, separating two
identical media. The eigenstates bounded to this interface, that we shall
call paired states, are then given by

\begin{eqnarray}  \label{bound}
f(l-m)=A e^{-\alpha |l-m|} \\
E= {\rm sgn}(U)\sqrt{U^2 + 16 \lambda^2 \cos^2 (ka)}  \nonumber
\end{eqnarray}

\noindent with $A$ a normalizing constant, and $\alpha =-{\rm arcsinh}%
(U/4\lambda \cos (ka))$. The range $\pi /2<ka<3\pi /2$ defines an energy
band covering the interval $U\leq E\leq \sqrt{U^{2}+16\lambda ^{2}}$ for $U>0
$, while a symmetric band of negative energies appears for $U<0$ and $-\pi
/2<ka<\pi /2$. Note that the interaction gives more weight to configurations
for which the electrons lie one on top of the other ($l\approx m$),
regardless of the sign of the interaction. Notice also that $\alpha $
diverges at the lower edge of the $U>0$ band (upper edge of the $U<0$ band),
leading to extreme localization with finite amplitudes along the interface $%
l=m$ only. The wavefunction is symmetric under exchange of particles and is
therefore appropriate to a singlet state. Together with this band of paired
states there is also a band of pure plane-wave solutions covering the
interval ($-4\lambda ,4\lambda $) which correspond to traveling waves that
scatter off the defect line $l=m$. The two bands giving paired and extended
states overlap except for $U>4\lambda $.

In discussing transport we consider the time dependent average position $%
<z>=\sum_{l,m}P(l,m)(l+m)a$ in the linear chain, where $P(l,m)=C^{\ast
}_{l,m}C_{l,m}$ is the probability of finding one electron at site $l$ and
the other at site $m$. We then take the time derivative of this expression
to find the velocity. After some algebraic manipulation we arrive with the
use of Eq. (1) at the following expression for the average velocity,
rigorous for the infinite chain,

\begin{equation}
<v>=-\frac{2\lambda a}{\hbar }Im%
\sum_{l,m}C_{l,m}(C_{l+1,m}^{*}+C_{l,m+1}^{*}).
\end{equation}

Exact results may be obtained in special cases with no disorder. For the
paired states (3) in the absence of an external field the average velocity
along the chain is a constant,

\begin{equation}
<v>=\frac{2\lambda a}{\hbar }\frac{\sin (ka)}{\cosh \alpha }.
\end{equation}

\noindent Note that the velocity is reduced by the interaction through the
denominator in this expression. In the presence of the field but this time
with no interaction the result is

\begin{equation}
<v>=\frac{2\lambda a}{\hbar }\sin (ka+eFat/\hbar ),
\end{equation}

\noindent giving the well known Bloch oscillations of period $T_{B}=h/eFa$.%
\cite{hack,bouc,clar} The combined case with external field and interactions
may be solved for $\lambda \ll eFa$ by noting that in this limit

\begin{equation}
C_{l,m}(t)=C_{l,m}(0)e^{-\frac{i}{\hbar }E_{l,m}t}+O(\lambda )\mbox{ . }
\label{aprox}
\end{equation}

\noindent so that owing to the presence of the external field, to the lowest
order of approximation the sites in the 2D lattice acquire different time
dependent phases. From Eqs. (3), (4) and (7) one then obtains for the paired
states the result, to $O(\lambda )$

\begin{equation}
<v>=\frac{2\lambda a}{\hbar }\sin (ka+\frac{eFat}{\hbar })\frac{\left(
1-2(1-e^{-2\alpha })\sin ^{2}\frac{Ut}{2\hbar }\right) }{\cosh \alpha }%
\mbox{ . }
\end{equation}

\noindent Note that the drift velocity is again decreased by the interaction
through the denominator in this expression, tending to zero as one
approaches the bottom of the band ($ka=\pi /2$ for $U>0$, $ka=0$ for $U<0$).
The result also shows that the coupling introduces an oscillation of period $%
T_{I}=h/U$. Although the amplitude of this interaction induced oscillation
(ININO) depends on the hopping amplitude $\lambda $, its period is
independent of this quantity and depends only on the interaction strength $U$%
.

A more general result for any form of the interaction potential may be
obtained if one assumes the system is in a plane wave (Bloch) state at $t=0$%
. Using Eqs. (4) and (7) and ignoring disorder one then gets, to $O(\lambda
) $

\begin{equation}
<v>=\frac{2\lambda a}{\hbar }\sin (ka+\frac{eFat}{\hbar })\frac{1}{N^{2}}%
\sum_{l,m}\cos (\frac{\delta V_{l,m}t}{\hbar })\mbox{ . }
\end{equation}

\noindent Here N is the number of sites and $\delta V_{l,m}=V(l+1-m)-V(l-m)$%
. The sum in this expression is bounded from above to $N^{2}$ adding
evidence that the interaction in general reduces the drift velocity. Also,
for the contact interaction model only $\delta V_{l,m}=U$ occurs in the
argument of the cosine when finite, so that, as in Eq. (8), there is an
oscillation of period $T_{I}=h/U$. For a long-range interaction several
frequencies may be present, however.

Next we consider disorder. From Eqs. (4) and (7) it is easy to see that the
contribution of a disordered distribution of site energies appears through
phase factors of the form $\exp [i(\epsilon _{l+1}-\epsilon _{l})t/\hbar ]$,
so that the average over disorder yields an overall factor $\ll \cos
[(\epsilon _{l+1}-\epsilon _{l})t/\hbar ]\gg $ in the drift velocity (4),
which at all times is less than one. Thus, disorder decreases the drift
velocity of the pair without affecting the period of the interaction-induced
oscillation discussed above.

Up to now our results rely on the approximation (7) that holds when $\lambda
\ll eFa$. We have performed numerical calculations to test all ranges of
parameters. A sample of our results are shown in Figs. (1) and (2). The
units of distance and time are $a$ and $\hbar /\lambda $, respectively. In
Fig. (1) we plot the time evolution of the center of mass drift velocity for 
$eFa=4\lambda $ without (Fig. 1a) and with (Fig. 1b) disorder, the latter
included through a random diagonal energy distribution in the interval $%
-5\lambda <\epsilon _{l}<5\lambda $. The solid line represents the Bloch
oscillation with no electron-electron interaction, while the doted curve is
for the contact interaction model with $U=100\lambda $. The dashed line adds
to the same contact interaction a Coulomb tail $V_{0}/|l-m|$ with $V_{0}=U/4$%
. Note first that, as exhibited by Eq. (8), the interaction reduces the
velocity. Note also that the ININO are clearly exhibited. They have a
regular period, and as anticipated in the above discussion, more than one
frequency is present in the long range interaction model. A Fourier analysis
of the data a strong ININO component redshifted by a factor of about 0.8.
The initial conditions for this data were finite uniform amplitudes in the
square $-M<l\pm m<M$ with M=3, and zero amplitude elsewhere. The sample was
a square lattice with up to 150 sites on each side, enough to avoid
significant reflections from the edges within the time of computation.
Increasing the size of M alters the relative amplitude of the oscillations
without modifying the period. Note that disorder does not destroy entirely
the ININO although there is an overall reduction in the velocity that
becomes more severe as time progresses.

Figure (2) shows the low field case $eFa=0.1\lambda $, M=10 and same value
of $U$ as above. Note that for these values of parameters in one Bloch
oscillation one expects a thousand ININO periods, only the first few of
which are shown. The almost perfectly straight (dashed) line is the
noninteracting result, bounding from above the correlated case obtained for
a uniform initial distribution, marked as $\alpha =0$. The curve labelled $%
\alpha =3.9116$ was obtained with an initial paired state as given by Eq.
(3) with $ka=0$. A different choice of $k$ just introduces a phase shift and
decreases the amplitude of the modulation, without afecting the period. The
results exhibited show that, as apparent from Eq. (8), for this rather large
value of the parameter $\alpha $ the state is dominated by the ININO and
motion is relatively slow. In this figure we chose to display a case with
small external field in order to illustrate our finding that the ININO exist
away from the limit in which Eqs. (8) and (9) hold as well. Note that in
spite of the diversity of initial conditions tested the oscillations are
always present.

The above results are for two electrons. The spectrum then includes two
relevant bands, one of extended states and one of bound states an energy U
away, in which the electrons tend to be on top of each other as described by
Eq. (3). The general case of N electrons may be treated in a similar way as
we did por two particles, resorting now to the equivalent problem of a
single electron in N dimensions with planar interfaces representing the
interaction. For instance, if N=3 one treats an electron in three dimensions
m,l,n, with uniform defect sheets along the planes $m=l,$ $m=n,$ $n=l$ and
an electric field along the diagonal $m=l=n$ where the defect planes meet.
Besides the band of extended states there are now two additional bands, one
an energy $\sim $U away, associated with interface states (our so-called
paired states), the other coming from states bound to the diagonal, an
energy $\sim $2U away (which we call tripled states). Figure 3 (a) shows the
center of mass velocity for this case using the same parameters as Fig. 1(a)
with contact interaction only. Initially, the amplitudes are set finite only
within a cube of side M=3 around the origin (0,0,0). The ININO oscillations
are clearly present, and as shown in Fig. 3(b) where the spectral density is
exhibited, include three main frequencies: $eFa/\hbar ,$ $U/\hbar ,$ and $%
2U/\hbar ,$ representing the Bloch oscillations and transitions between the
three bands. Notice that the weakest frequency is for oscillations involving
the highest band. This is an important consequence of the reduced number of
states in the interaction-induced upper bands. Figure 3(c) shows the
spectral density for finite initial amplitudes over a similar cube as for
Fig 3(b) but sorrounding the point (0,0,20). Notice that this point is far
from the diagonal $m=l=n$ near which tripled states are localized so that no
component in the highest band is expected. Indeed, the spectral density of
the highest frequency $2U/\hbar $ is negligible as is apparent in the
figure. In the general case of N particles in a string of L sites the number
of extended states equal about L$^{N}$ while the paired states number L$%
^{N-1}$, the tripled states L$^{N-2}$, the cuadrupleds L$^{N-3},$ and so on$%
. $ Thus the amplitude of their contribution decreases in the same ratio and
only the lowest of such bands is important. Then, a system with more than
two particles will still exhibit the oscillations described. The equivalence
of the 1D wire with N electrons and the motion of a single particle in N
dimensions emphasized above may also be usefull in checking the effect we
are reporting. For instance, the experimental probe could be either a 1D
system with three electrons, or one electron moving in a 3D lattice hosting
a thin sheet of impurities.

In summary, we have shown that the electron-electron interaction induces a
new kind of oscillations in the drift velocity of electrons moving along a
chain and subject to an external electric field, with a period determined
solely by the interaction range and strength. The N particle problem is
identical to that of a single particle moving in an N-dimensional lattice,
with defect surfaces dividing the space in symmetric domains. One can take
advantage of the equivalence of the two cases to understand the physical
origin of the oscillations. With no interactions the single particle in N
dimensions will respond to an external field purely through Bloch
oscillations in a band of extended states, Eq. (6). The defect boundary
introduced by the coupling gives rise to separate bands of surface states
localized along the line perpendicular to the surface, such that electrons
may exhibit oscillations between the free- and bound-states bands. This
interpretation is supported by our numerical results showing that the ININO
dissapear if one starts with a state with finite amplitudes far from the
defect line only. In fact, as the square around the origin in which
amplitudes are initially finite in Fig. 2 is enlarged, the oscillation of
the upper curve is flattened due to the larger component in the lower
extended-states energy band of the initial state, while the oscillation in
the lower curve remains. If $U$ is negative so that the band of paired
states becomes the lowest in energy, then this latter oscillation is the one
damped out.

This research was carried out with support of a C\'{a}tedra Presidencial en
Ciencias (F.C.) and FONDECYT, Grants 1020829 and 1010776.

\newpage

{\bf {Figure captions}} \vskip 10pt

\begin{itemize}
\item[ ]  {Fig. 1.} Center of mass drift velocity for a pair of electrons in
an electric field, without (a) and with (b) disorder. The dotted (full) line
is the evolution with (without) a contact interaction. The dashed line
includes a contact potential as well as a Coulomb tail. For details see text.

\item[ ]  {Fig. 2.} Drift velocity for an interacting pair initially in a
finite square on the $l,m$ plane, with 11 sites on the side (full line
labelled $\alpha =0)$. The case $\alpha =3.9116$ has initial amplitudes as
given by Eq. (3). The dashed line is the non-interacting case

Fig. 3. Drift velocity (a), and spectral density for three interacting
particles with finite initial amplitudes around the origin (b) and around
the point (0,0,20) (c). Parameters are as in Fig. 1(a).
\end{itemize}

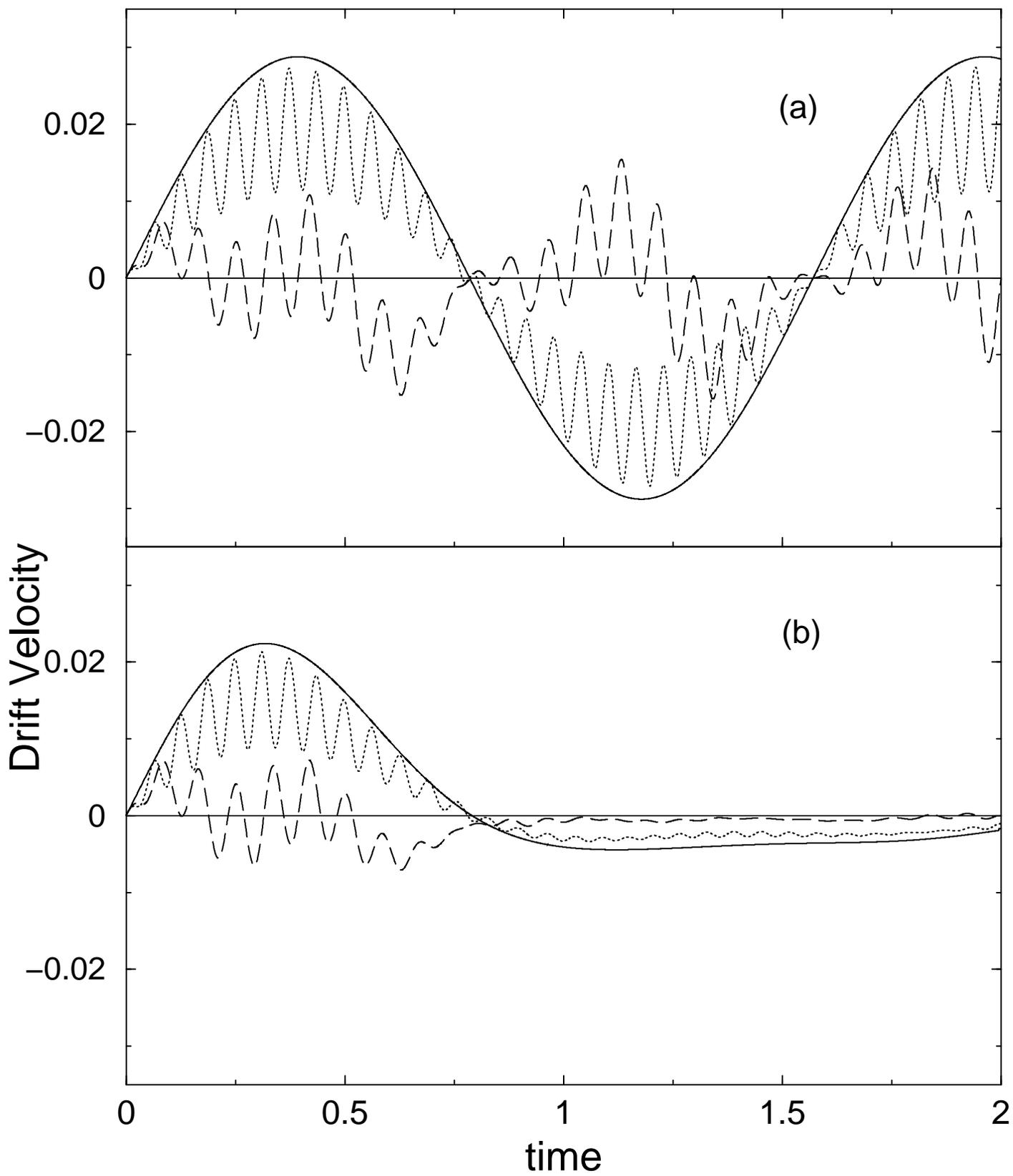
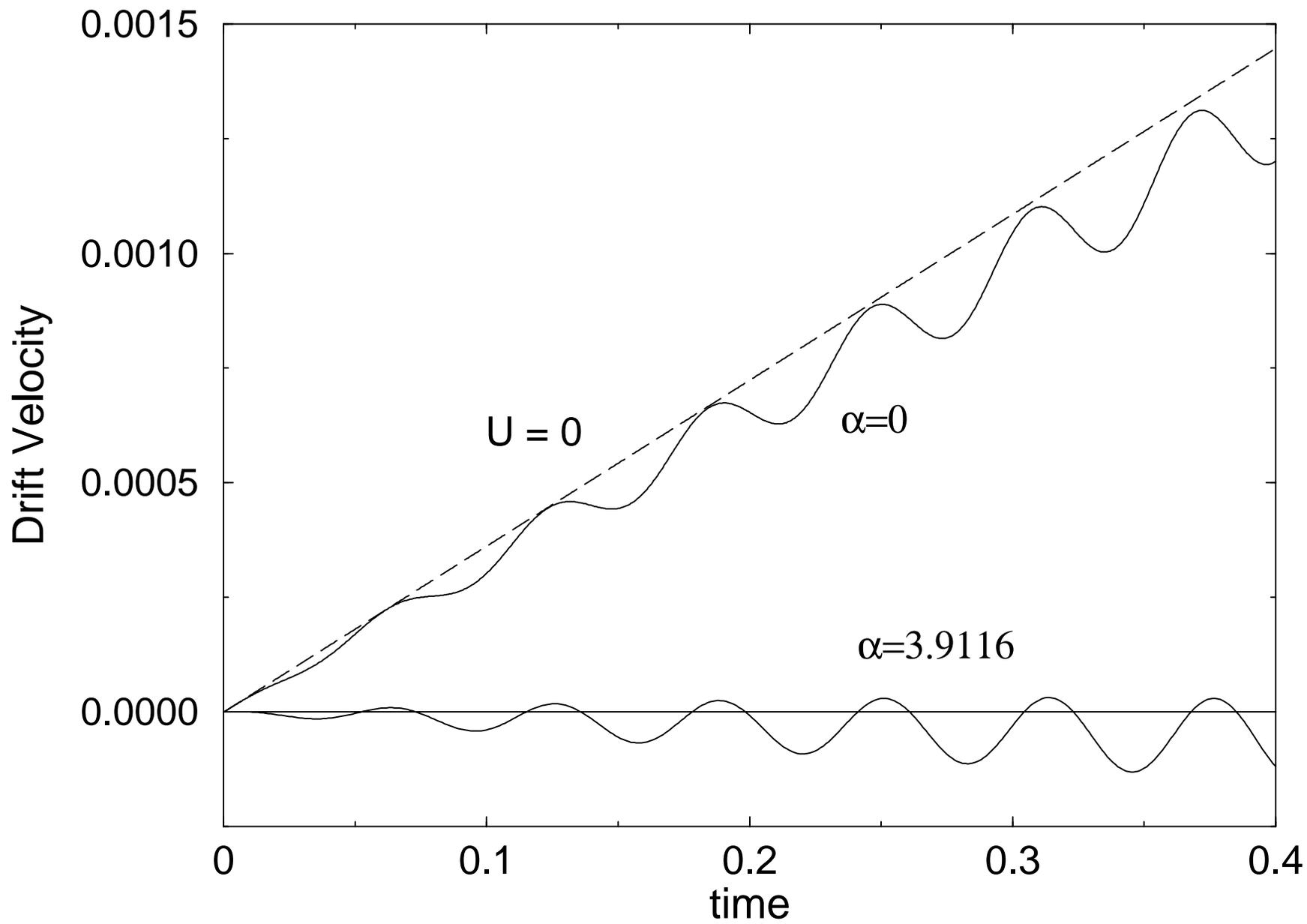
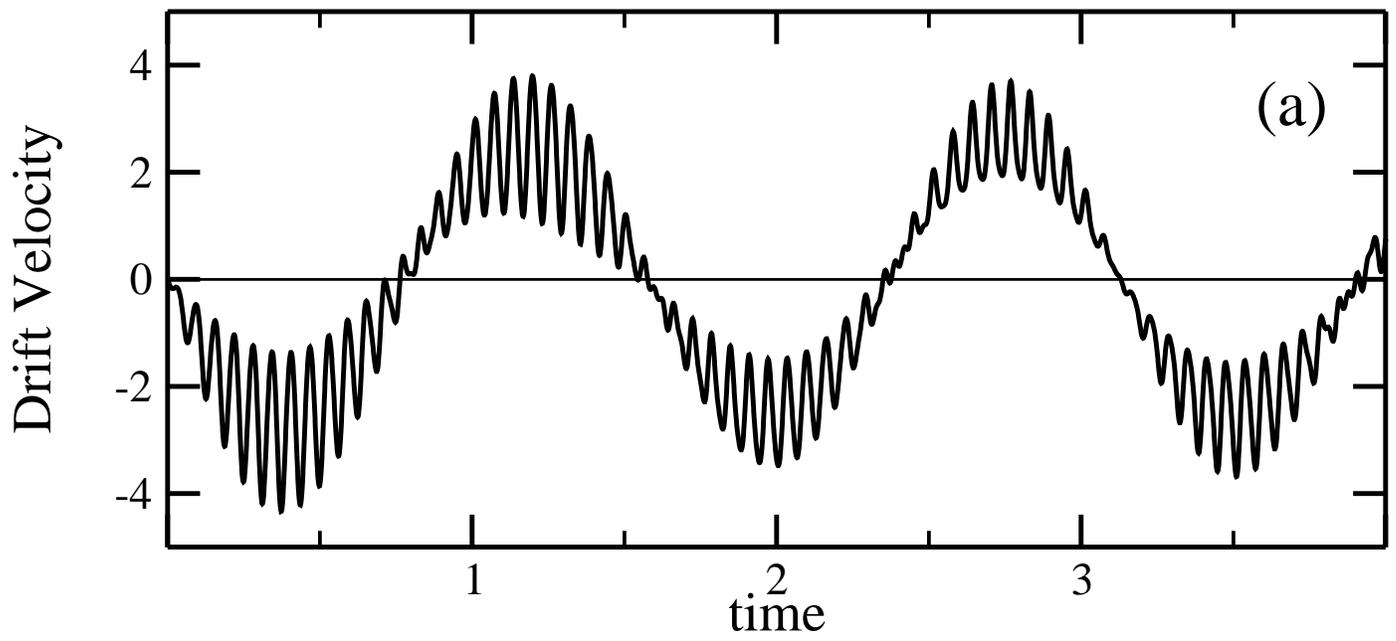
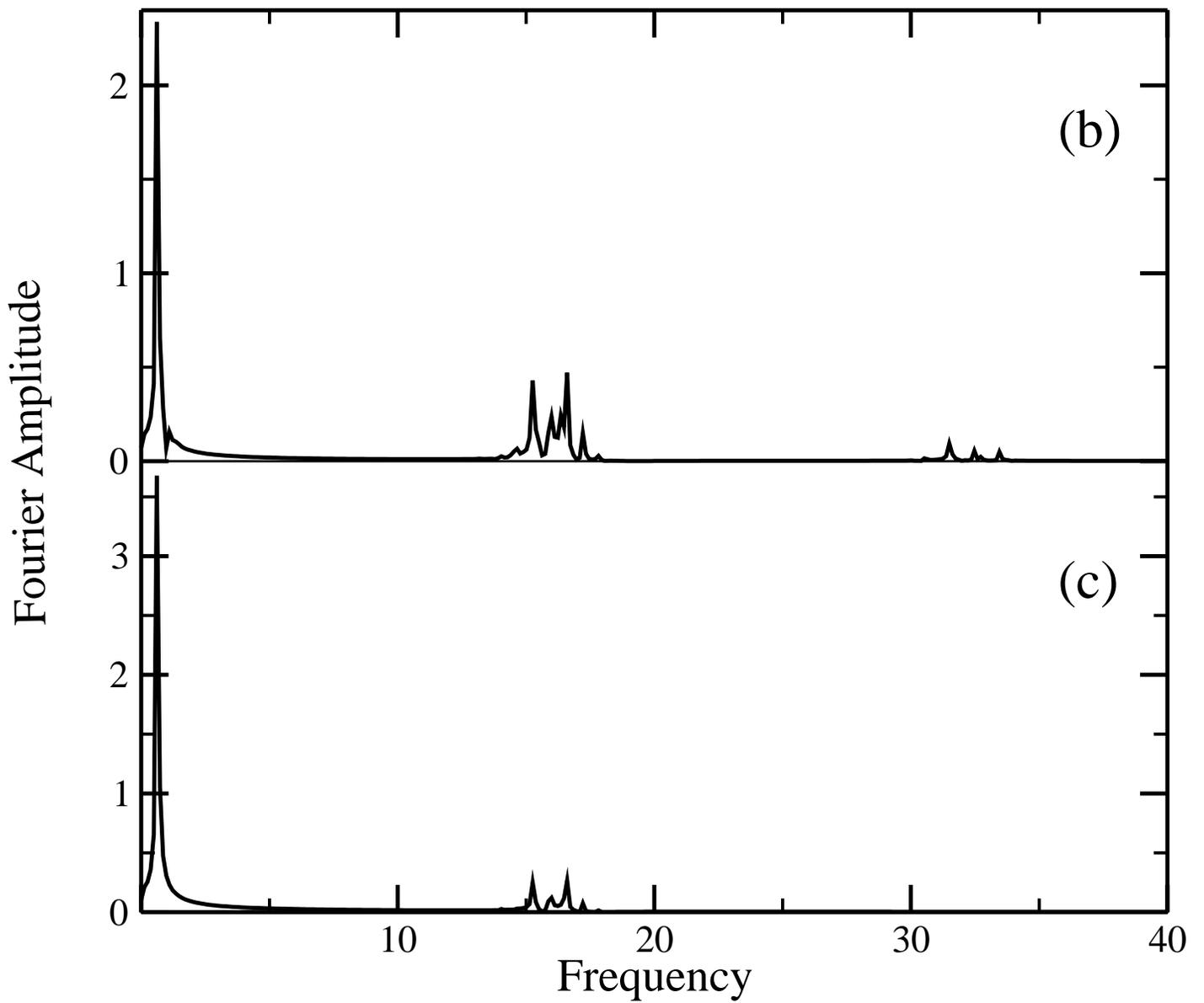
\end{document}